\documentclass[ floatfix,twocolumn]{revtex4}
\usepackage{amsmath}
\usepackage{color}
\usepackage{amssymb}

\usepackage{graphicx}

\begin{document}

\title{ Electron-Electron Scattering  and Resistivity in Non-Centrosymmetric Metals}

\author{V.P.Mineev$^{1,2\footnote{E-mail: vladimir.mineev@cea.fr}}$}
\affiliation{$^1$Universite Grenoble Alpes, CEA, IRIG, PHELIQS, F-38000 Grenoble, France\\
$^2$Landau Institute for Theoretical Physics, 142432 Chernogolovka, Russia}

\begin{abstract}
The quadratic low-temperature dependence of resistance in ordinary metals is determined by the momentum relaxation  due to electron-electron scattering
in the presence  Umklapp processes and scattering on impurities. In metals without inversion center spin-orbit interaction of electrons with crystal lattice  lifts spin degeneracy of electron states and splits each band on two bands. 
The temperature dependence of electron-electron collisions scattering rate is found using  the matrix kinetic  equation including the electron-impurity and electron-electron intra and inter band scattering.
It is shown that in clean enough case when 
the energy of band splitting exceeds  the electron-impurity scattering rate but at the same time it is much smaller than the Fermi energy the square low-temperature 
dependence of resistivity is still valid.
\end{abstract}

\date{\today}
\maketitle

\section{Introduction}
The low temperature dependence of normal metals resistivity 
\begin{equation}
\rho=\rho_0+AT^2
\label{1}
\end{equation}
is determined by electron-electron scattering including Umklapp processes with scattering rate \cite{Landau1936,Abrikosov} 
\begin{equation}
\frac{1}{\tau_{ee}}\propto \frac{V^2}{\varepsilon_F^2}\frac{T^2}{\varepsilon_F},
\label{2}
\end{equation}
where $V$ is the amplitude of screened short range potential of electron-electron interaction, 
 $\varepsilon_F$ is the Fermi energy.
Usually $1/\tau_{ee}$ is quite small and the dependence given by Eq.(\ref{1}) is observable in metals with narrow conducting bands or in heavy fermion compounds with small energy Fermi. 
 In absence of Umklapp processes in single band metals and  semiconductors the contribution from the electron-electron scattering play no role and the stationary conductivity is determined only by the scattering on impurities \cite{Keyes1958}. In a multiband metal, however, the contribution from the interband electron-electron scattering survives \cite{Gantmakher,Appel1978,Murzin,Pal2012} and determines the temperature dependence of resistivity given by Eq.(1).

The square temperature dependence arises because  according to the Pauli principle the electrons can scatter each other only in a narrow energy layer of the order of temperature near the Fermi surface.
This property takes place also in metals with several conducting bands.

In metals without inversion center the spin-orbit interaction splits the 
the Fermi surface in each conducting band to two Fermi surfaces with different Fermi momenta. 
In the paper \cite{Mineev2018} there was derived 
 the following
   scattering rate  of momentum relaxation   due to electron-electron collisions 
\begin{equation}
\frac{1}{\tau_{ee}}\propto\frac{(2\pi T)^2+(v_F\Delta k_F)^2}{\varepsilon_F},
\label{11}
\end{equation}
where $\Delta k_F=k_{F+}-k_{F-}$. This expression actually means that in metals without an inversion center the residual resistance at T = 0
greatly increases due to splitting of zones by spin-orbital interaction.
 However,  the derivation presented in \cite{Mineev2018}.  was done making use
the  electron-electron scattering integral in form inapplicable to metals without space parity.
The proper collision integral was derived in the following paper \cite{Mineev2019}.  
The corresponding calculations 
confirmed the expression (\ref{11}). In the  both  calculations there were used  the dispersion laws $ \xi_{\pm}(k)=v_F(k-k_{F\pm})$ for electrons in two energy bands split by the spin-orbit interaction. Each of these expressions is valid near the corresponding Fermi surface with radii $k_{F+}$ and  $k_{F-}$. But $\xi_+(k)$ is wrong near the
Fermi surface with the Fermi momentum $k_{F-}$ as well as  $\xi_-(k)$ is wrong near the
Fermi surface with the Fermi momentum $k_{F+}$. This error has been fixed in the present work.

 An analytic derivation of dependence given by Eq. (\ref{2}) in case of arbitrary shape of the Fermi surface is not possible even for a metal with  a single conducting  band in absence of space parity violation. The point is that the length of the Fermi momentum is varied from point to point at the Fermi surface. All known calculations were made under the implicit assumption that the variation of the Fermi momentum length is much smaller than its average value. In addition to this in our calculation we also assume the smallness of the band splitting energy in comparison with the energy Fermi:
 \begin{equation}
\varepsilon_F\gg v_F\Delta k_F.
\label{4}
\end{equation}

The kinetics of a metal without inversion center is described by kinetic equation for a matrix electron distribution function.
The neglect of its off-diagonal (inter-band) elements  is possible at assumption
  that the energy of band splitting exceeds  the  scattering rate of electrons on impurities,  on electrons, etc.
The electron-electron scattering rate is always small in comparison with  the electron scattering rate on impurities $1/\tau_i$. Thus, we will assume
\begin{equation}
v_F\Delta k_F\gg\frac{1}{\tau_{i}}.
\label{5}
\end{equation}

The paper is organised as follows.
 In the next section there are given the kinetic equations for the electron gas in noncentrosymmetric metals.
The calculations
presented  in the third section show that in   the assumptions (\ref{4}) and (\ref{5}) 
 the low temperature dependence of momentum relaxation rate in metals without inversion center is given by the Eq.(2).

\section{Kinetic equation in non-centrosymmetric metals}

 The spectrum of noninteracting
electrons in a metal without inversion center is:
\begin{equation}
\label{H_0}
 \hat \varepsilon({\bf k})
 = \varepsilon({\bf k})\sigma_0+\mbox{\boldmath$\gamma$}({\bf k}) 
   \cdot \mbox{\boldmath$\sigma$},
\end{equation}
where 
$\varepsilon({\bf k})$ denotes the spin-independent part of the spectrum,
$\sigma_0$ is the unit $2\times 2$ matrix in the spin space, $\mbox{\boldmath$\sigma$}=(\sigma_x,\sigma_y,\sigma_z)$ are the Pauli matrices. 
The second term in Eq.
(\ref{H_0}) describes the  spin-orbit  coupling whose form depends on the specific noncentrosymmetric crystal structure.
The pseudovector $\mbox{\boldmath$\gamma$}({\bf k})$  is periodic 
 in the reciprocal space function and         satisfies
$\mbox{\boldmath$\gamma$}(-{\bf k})=-\mbox{\boldmath$\gamma$}({\bf k})$ and 
$g\mbox{\boldmath$\gamma$}(g^{-1} {\bf k})=\mbox{\boldmath$\gamma$}({\bf k})$,
where $g$ is any symmetry operation in the  point group ${\cal G}$ of
the crystal. Near the $\Gamma$ point , in the case of cubic symmetry   
\begin{equation}
\label{gammaO}
    \mbox{\boldmath$\gamma$}({\bf k})=\gamma{\bf k}.
\end{equation}
 Here $\gamma$ is a constant. 
In  case of the tetragonal point group
${\cal G}=\mathbf{C}_{4v}$
 the antisymmetric spin-orbit coupling is
\begin{equation}
\label{gammaC4v}
\mbox{\boldmath$\gamma$}({\bf k})=\gamma(k_y\hat x-k_x\hat y)
    +\gamma_\parallel k_xk_yk_z(k_x^2-k_y^2)\hat z.
\end{equation}
In the purely two-dimensional case, setting $\gamma_\parallel=0$
one recovers the Rashba interaction \cite{Rashba1960}.

The eigenvalues and the eigenfunctions of the matrix (\ref{H_0}) are
\begin{equation}
    \varepsilon_{\pm}({\bf k})=\varepsilon({\bf k})\pm|\mbox{\boldmath$\gamma$}({\bf k})|,
\label{e3}
\end{equation}
\begin{eqnarray}
\Psi^+_\sigma({\bf k})=C_{\bf k}\left (\begin{array} {c}
\hat\gamma_{{\bf k}z}+1\\
\hat\gamma_{{\bf k}x}+i\hat\gamma_{{\bf k}y}
\end{array}\right),\nonumber\\
~~~~~~~~~~~~\Psi^-_\sigma({\bf k})=C_{\bf k}\left(\begin{array} {c}
-\hat\gamma_{{\bf k}x}+i\hat\gamma_{{\bf k}y}\\
\hat\gamma_{{\bf k}z}+1
\end{array}\right),\\
~~~~~~~~~~C_{\bf k}=(2(\gamma_{{\bf k}z}+1))^{-1/2},\nonumber\\
\hat\gamma_{\bf k}=\frac{\mbox{\boldmath$\gamma$}({\bf k}) }{|\mbox{\boldmath$\gamma$}({\bf k}) |}.~~~~~~~~~~~~~\nonumber
\end{eqnarray}
 The eigen functions obey the orthogonality conditions
\begin{equation}
\Psi^{\alpha\star}_\sigma({\bf k})\Psi^\beta_\sigma({\bf k})=\delta_{\alpha\beta},~~~~~~~
\Psi^\alpha_{\sigma_1}({\bf k})\Psi^{\alpha\star}_{\sigma_2}({\bf k})=\delta_{\sigma_1\sigma_2}.
\label{ort}
\end{equation}
Here, and in all the subsequent formulas there is implied the summation over the repeating  spin $\sigma=\uparrow,\downarrow$
or band $\alpha=+,-$ indices.

There are two Fermi surfaces with different Fermi momenta ${\bf k}_{F\pm}$ determined by the equations
\begin{equation}
\label{e4}
    \varepsilon_{\pm}({\bf k})=\mu.
\end{equation}
In the Rashba 2D model and in the 3D isotropic case they are
\begin{equation}
k_{F\pm}=\mp m \gamma+\sqrt{2m\mu+(m\gamma)^2}
\end{equation}
and the Fermi velocity  has the  common value 
\begin{equation}
{\bf v}_{F\pm}=\frac{\partial(\varepsilon_{\pm}({\bf k})}{\partial {\bf k}}|_{k=k_{F\pm}}=\hat{\bf k}\sqrt{\frac{2\mu}{m}+\gamma^2},
\end{equation}
here $\hat{\bf k}$ is the unit vector along momentum ${\bf k}$. The equivalence of the Fermi velocities at different Fermi momenta is the particular property of the models with isotropic spin-orbital coupling (\ref{gammaO}) in 3D case and the Rashba interaction in 2D case.

The matrix of equilibrium electron distribution function is
\begin{equation}
\hat n
=\frac{n({\varepsilon}_+)+n({\varepsilon}_-)}{2}\sigma_0+\frac{n({\varepsilon}_+)-n({\varepsilon}_-)}{2|\mbox{\boldmath$\gamma$}|} \mbox{\boldmath$\gamma$} \cdot \mbox{\boldmath$\sigma$},
\label{eqv}
\end{equation}
where 
\begin{equation}
n({\varepsilon})=\frac{1}{\exp\left(\frac{\varepsilon-\mu}{T}\right)+1}
\end{equation}
is the Fermi function.

The hermitian matrices of the nonequilibrium distribution functions in band and spin representations are related as 
\begin{equation}
f_{\alpha\beta}({\bf k})=\Psi^{\alpha\star}_{\sigma_1}({\bf k})n_{\sigma_1\sigma_2}\Psi^{\beta}_{\sigma_2}({\bf k}).
\label{f}
\end{equation}
In the band representation the equilibrium distribution function  is the diagonal matrix
\begin{equation}
n_{\alpha\beta}=\Psi^{\alpha\star}_{\sigma_1}({\bf k})n_{\sigma_1\sigma_2}\Psi^{\beta}_{\sigma_2}({\bf k})=\left (\begin{array} {cc} n({\varepsilon}_+)&0\\0&n({\varepsilon}_-)  \end{array}\right)_{\alpha\beta}.
\end{equation}
So, the off-diagonal matrix elements $f_{+-}({\bf k},~f_{-+}({\bf k}$ are not equal to zero only out equilibrium.

The general form of kinetic equation for the electron distribution function in non-centrosymmetric metals is derived in the paper \cite{Mineev2019}. 
The corresponding stationary condition 
 in external electric field  ${\bf E}$
 is
\begin{widetext}
\begin{eqnarray}
e\left (\begin{array} {cc}({\bf v}_{+}{\bf E}) \frac{\partial n({\varepsilon}_+)}{\partial \varepsilon_+}&({\bf v}_{\pm}{\bf E})(n({\varepsilon}_-)-n({\varepsilon}_+))
\\
({\bf v}_{\mp}{\bf E})(n({\varepsilon}_+)-n({\varepsilon_-}))&({\bf v}_{-}{\bf E})\frac{\partial n({\varepsilon}_-)}{\partial \varepsilon_-}
 \end{array}\right)+
 \left(
\begin{array} {cc}0&i(\varepsilon_--\varepsilon_+)f_{\pm}({\bf k})\\
i(\varepsilon_+-\varepsilon_-)f_{\mp}({\bf k})&0
 \end{array}\right)=\hat I^{i}+\hat I^{ee}
 \label{eqv1}
\end{eqnarray}
Here 
\begin{eqnarray}
{\bf v}_\alpha({\bf k})=\frac{\partial\varepsilon_\alpha}{\partial{\bf k}},~~~~
{\bf v}_{\pm}({\bf k})=
\Psi^{+\star}_{\sigma}({\bf k})\frac{\partial \Psi^{-}_{\sigma}({\bf k})}{\partial{\bf k}},~~~
{\bf v}_{\mp}=-{\bf v}_{\pm}^\star.
\label{vel}
\end{eqnarray}
In Born approximation the collision integral $I^{i}_{\alpha\beta}$ for electron scattering on impurities is 
\begin{eqnarray}
I^i_{\alpha\beta}({\bf k})=2\pi n_{imp}\int\frac{d^3k^\prime}{(2\pi)^3}|V({\bf k}-{\bf k}^\prime)|^2\left \{O_{\alpha\nu}({\bf k},{\bf k}^\prime)\left [ f_{ \nu\mu}({\bf k}^\prime)O_{\mu\beta}({\bf k}^\prime,{\bf k})-O_{\nu\mu}({\bf k}^\prime,{\bf k})
  f_{ \mu\beta}({\bf k}) \right ]\delta(\varepsilon^\prime_\nu-\varepsilon_\beta)\right.\nonumber\\ 
 \left .+
  \left[O_{\alpha\nu}({\bf k},{\bf k}^\prime)f_{ \nu\mu}({\bf k}^\prime)-f_{ \alpha\nu}({\bf k})O_{\nu\mu}({\bf k},{\bf k}^\prime)
  \right ]O_{\mu\beta}({\bf k}^\prime,{\bf k})\delta(\varepsilon^\prime_\mu-\varepsilon_\alpha)\right \}.
  \label{matrix1}
\end{eqnarray}
Here, we introduced notations $\varepsilon_{\alpha}=\varepsilon_{\alpha}({\bf k}),~ \varepsilon_{\mu}^\prime=\varepsilon_{\mu}({\bf k}^\prime)$ etc,
\begin{equation}
O_{\alpha\beta}({\bf k},{\bf k}^\prime)=\Psi^{\alpha\star}_\sigma({\bf k})\Psi^\beta_\sigma({\bf k}^\prime)
\end{equation}
such that
$
O_{\alpha\beta}({\bf k},{\bf k}^\prime)=O^\star_{\beta\alpha}({\bf k}^\prime,{\bf k}).
$
The  electron-electron  collisions  integral in the Born approximation \cite{Mineev2019}
  is:  
\begin{equation}
\hat I^{ee}({\bf k})=2\pi\int \frac{d^3{\bf k}^{\prime\prime}}{(2\pi)^3}\frac{d^3{\bf k}_2}{(2\pi)^3}
 \hat F({\bf k},{\bf k}_{2},{\bf k}^\prime,{\bf k}^{\prime\prime}),
\label{e27}
\end{equation}
where $ {\bf k}^\prime={\bf k}+{\bf k}_{2}-{\bf k}^{\prime\prime}-{\bf Q}$  and ${\bf Q}$ is a vector of reciprocal lattice. Throughout the paper we put the Planck constant $\hbar=1$.
The matrix $\hat F$
 is
\begin{eqnarray}
F_{\alpha\beta}({\bf k},{\bf k}_{2},{\bf k}^\prime,{\bf k}^{\prime\prime})=~~~~~~~~~~~~~~~~~~~~~~~~~~~~~~~~~~~~~~~~~~~~~~~~~~~~~~~~\nonumber\\
=\frac{1}{2}W_1\left \{  \left [O_{\alpha\nu}({\bf k},{\bf k}^\prime) f_{ \nu\mu}({\bf k}^\prime)O_{\mu\lambda}({\bf k}^\prime,{\bf k})(\delta_{\lambda\beta}-f_{\lambda\beta}({\bf k}))~(\delta_{\xi\eta}-f_{\xi\eta}({\bf k}_2))O_{\eta\zeta}({\bf k}_2,{\bf k}^{\prime\prime}) f_{\zeta\rho}({\bf k}^{\prime\prime})O_{\rho\xi}({\bf k}^{\prime\prime},{\bf k}_2)\right.\right.~~~~~~~~~~~~~~~~~~~~\nonumber\\
~\nonumber\\
-\left.\left.
O_{\alpha\nu}({\bf k},{\bf k}^\prime)(\delta_{\nu\mu}- f_{ \nu\mu}({\bf k}^\prime))O_{\mu\lambda}({\bf k}^\prime,{\bf k})
f_{\lambda\beta}({\bf k})~
f_{\xi\eta}({\bf k}_2)O_{\eta\zeta}({\bf k}_2,{\bf k}^{\prime\prime})(\delta_{\zeta\rho}- f_{\zeta\rho}({\bf k}^{\prime\prime}))O_{\rho\xi}({\bf k}^{\prime\prime},{\bf k}_2)\right ] 
\right.\delta(\varepsilon_\nu^\prime-\varepsilon_{\beta} -\varepsilon_{2\xi} +\varepsilon_\zeta^{\prime\prime}) \nonumber\\
~\nonumber\\
+
  \left.\left [(\delta_{\alpha\nu}-f_{\alpha\nu}({\bf k}))O_{\nu\mu}({\bf k},{\bf k}^\prime)f_{\mu\lambda}({\bf k}^\prime)O_{\lambda\beta}({\bf k}^\prime,{\bf k})~
  (\delta_{\xi\eta}-f_{\xi\eta}({\bf k}_2))O_{\eta\zeta}({\bf k}_2,{\bf k}^{\prime\prime}) f_{\zeta\rho}({\bf k}^{\prime\prime})O_{\rho\xi}({\bf k}^{\prime\prime},{\bf k}_2)\right.\right.~~~~~~~~~~~~~~~~~~~~ \nonumber\\
 ~\nonumber\\
 -
\left.\left.f_{\alpha\nu}({\bf k}))O_{\nu\mu}({\bf k},{\bf k}^\prime)
(\delta_{\mu\lambda}-f_{\mu\lambda}({\bf k}^\prime))O_{\lambda\beta}({\bf k}^\prime,{\bf k})
f_{\xi\eta}({\bf k}_2)O_{\eta\zeta}({\bf k}_2,{\bf k}^{\prime\prime})(\delta_{\zeta\rho}- f_{\zeta\rho}({\bf k}^{\prime\prime}))O_{\rho\xi}({\bf k}^{\prime\prime},{\bf k}_2)\right ]\delta(\varepsilon_{\alpha}-\varepsilon_\mu^{\prime}+\varepsilon_{2\xi} -\varepsilon_{\zeta}^{\prime\prime} )\right\} \nonumber\\
~\nonumber\\
+ 
\frac{1}{2}W_2\left \{  \left [O_{\alpha\nu}({\bf k},{\bf k}^\prime) f_{ \nu\mu}({\bf k}^\prime)O_{\mu\lambda}({\bf k}^\prime,{\bf k}_2)(\delta_{\lambda\xi}-f_{\lambda\xi}({\bf k}_2))O_{\xi\zeta}({\bf k}_2,{\bf k}^{\prime\prime})
f_{\zeta\rho}({\bf k}^{\prime\prime}))
O_{\rho\omega}({\bf k}^{\prime\prime},{\bf k})(\delta_{\omega\beta}- f_{\omega\beta}({\bf k})~~~~~~~~~~~~~~~~~~~~\right.\right.\nonumber\\
~\nonumber\\
-\left.\left.O_{\alpha\nu}({\bf k},{\bf k}^\prime)(\delta_{\nu\mu}- f_{ \nu\mu}({\bf k}^\prime))O_{\mu\lambda}({\bf k}^\prime,{\bf k}_2)f_{\lambda\xi}({\bf k}_2)O_{\xi\zeta}({\bf k}_2,{\bf k}^{\prime\prime})
(\delta_{\zeta\rho}-f_{\zeta\rho}({\bf k}^{\prime\prime}))
O_{\rho\omega}({\bf k}^{\prime\prime},{\bf k})f_{\omega\beta}({\bf k}\right ] \delta(\varepsilon_\nu^\prime-\varepsilon_{\beta} -\varepsilon_{2\xi} +\varepsilon_\zeta^{\prime\prime})\right.\nonumber\\
~\nonumber\\
+ 
\left.  \left [(\delta_{\alpha\nu}- f_{\alpha\nu}({\bf k})O_{\nu\mu}({\bf k},{\bf k}^\prime) f_{ \mu\lambda}({\bf k}^\prime)O_{\lambda\xi}({\bf k}^\prime,{\bf k}_2)(\delta_{\xi\zeta}-f_{\xi\zeta}({\bf k}_2))O_{\zeta\rho}({\bf k}_2,{\bf k}^{\prime\prime})
f_{\rho\omega}({\bf k}^{\prime\prime}))
O_{\omega\beta}({\bf k}^{\prime\prime},{\bf k})~~~~~~~~~~~~~~~~~~~~~~~~~\right.\right.\nonumber\\
~\nonumber\\
-
\left.  \left. f_{\alpha\nu}({\bf k})O_{\nu\mu}({\bf k},{\bf k}^\prime)(\delta_{\mu\lambda}- f_{ \mu\lambda}({\bf k}^\prime))
O_{\lambda\rho}({\bf k}^\prime,{\bf k}_2)f_{\rho\xi}({\bf k}_2)O_{\xi\zeta}({\bf k}_2,{\bf k}^{\prime\prime})
(\delta_{\zeta\omega}-f_{\zeta\omega}({\bf k}^{\prime\prime}))
O_{\omega\beta}({\bf k}^{\prime\prime},{\bf k})\right]\delta(\varepsilon_{\alpha}-\varepsilon_{\mu}^\prime +\varepsilon_{2\xi} -\varepsilon_\zeta^{\prime\prime})\right\}.
\nonumber\\
\label{eqv2}
\end{eqnarray}
\end{widetext}
Here, $W_1$,
$W_2$  are  the momenta dependent amplitudes of direct and exchange interaction correspondingly.
In concrete metal they are  unknown and due to  charge screening one can put them by the constants.

The off-diagonal terms in the left hand side of the matrix kinetic equation (\ref{eqv1}) are proportional to the band splitting energy
whereas the integral terms in the right hand side are proportional to the different intraband and interband electron-impurity and electron-electron  scattering rates.
The electron-electron scattering rate is always small in comparison with the electron scattering rate on impurities.
Hence, 
if the energy of band splitting exceeds the electron-
impurity scattering rate
\begin{equation}
v_F(k_{F-}-k_{F+})\gg1/\tau_{i}
\end{equation}
one can neglect by the collision integrals 
in the off-diagonal terms of matrix kinetic equation  (\ref{eqv1}) and use
the collision-less
solution  for the off-diagonal terms of the matrix distribution function 
\begin{eqnarray}
f_{\pm}=e({\bf w}_{\pm}{\bf E})=\frac{e({\bf v}_{\pm}{\bf E})(n_--n_+)}{i(\varepsilon_--\varepsilon_+)},
\label{C}\\
f_{\mp}=e({\bf w}_{\mp}{\bf E})=\frac{e({\bf v}_{\mp}{\bf E})(n_+-n_-)}{-i(\varepsilon_+-\varepsilon_-)}.
\label{D}
\end{eqnarray}
There was shown that  in stationary case this type of the off-diagonal terms do not produce  a contribution to the electric current (see Ref.9).
On the other hand, substitution of these expressions to the diagonal parts of collision-integral matrices  (\ref{matrix1}) and (\ref{eqv2})
allows to neglect in them by all the terms containing off-diagonal elements of distribution function. These terms are $\gamma k_F\tau_i>>1$ times smaller than the terms
with diagonal elements.
Then the system Eq.({\ref{eqv1}) for
\begin{equation}
f_{\alpha\beta}({\bf k})=
\left (\begin{array} {cc} f_+({\bf k})&0\\0&f_-({\bf k})  \end{array}\right)_{\alpha\beta}
\end{equation}
acquires the following form:
\begin{equation}
({\bf v}_{+}{\bf E}) \frac{\partial n({\varepsilon}_+)}{\partial \varepsilon_+}=I_+^i+I_+^{ee},
\label{k}
\end{equation}
\begin{equation}
({\bf v}_{-}{\bf E}) \frac{\partial n({\varepsilon}_-)}{\partial \varepsilon_-}=I_-^i+I_-^{ee},
\label{l}
\end{equation}
where
\begin{widetext}
\begin{eqnarray}
I_+^i=4\pi n_i\int\frac{d^3k}{2\pi^3}|V({\bf k}-{\bf k}^\prime)|^2\times\nonumber~~~~~~~~~~~~~~~~~~~~~~~~~~~~~~~~~~~~~~~~~~~~~~
\\ \times\left\{
O_{++}({\bf k}{\bf k}^\prime)O_{++}({\bf k}^\prime{\bf k})[f_+({\bf k}^\prime)-f_+({\bf k})]\delta(\varepsilon_+^\prime-\varepsilon_+)+
O_{+-}({\bf k}{\bf k}^\prime)O_{-+}({\bf k}^\prime{\bf k})[f_-({\bf k}^\prime)-f_+({\bf k}))]\delta(\varepsilon_-^\prime-\varepsilon_+)
\right \},~~\label{34}\\
I_+^{ee}=2\pi\int \frac{d^3{\bf k}^{\prime\prime}}{(2\pi)^3}\frac{d^3{\bf k}_2}{(2\pi)^3}\times~~~~~~~~~~~~~~~~~~~~~~~~~~~~~~~~~~~~~~~~~~~~~~~~~~~
\nonumber\\
\times \left\{W_1 \left [O_{+\nu}({\bf k}, {\bf k}^\prime)) O_{\nu +}({\bf k}^\prime, {\bf k}))
O_{\xi\zeta}({\bf k}_2, {\bf k}^{\prime\prime})
O_{\zeta\xi}({\bf k}^{\prime\prime}, {\bf k}_2)\right ]
+W_2\left   [O_{+\nu}({\bf k}, {\bf k}^\prime)) O_{\nu \xi}({\bf k}^\prime, {\bf k}_2))
O_{\xi\zeta}({\bf k}_2, {\bf k}^{\prime\prime})
O_{\zeta +}({\bf k}^{\prime\prime}, {\bf k}) \right ]\right\}\nonumber\\
\times \left\{  f_{\nu}({\bf k}^\prime)(1- f_{+}({\bf k}))(1-f_{\xi}({\bf k}_2))f_{\zeta}({\bf k}^{\prime\prime})-\right.
\left.(1- f_{\nu}({\bf k}^\prime)) f_{+}({\bf k})f_{\xi}({\bf k}_2)(1-f_{\zeta}({\bf k}^{\prime\prime}) )\right\}
\delta(\varepsilon_\nu^\prime-\varepsilon_{+} -\varepsilon_{2\xi} +\varepsilon_\zeta^{\prime\prime}).
\label{56}
\end{eqnarray}
\end{widetext}
The corresponding expressions for  $I_-^i$ and $I_-^{ee}$ are obtained by the interchange $(+\leftrightarrow -)$.
Thus, we came to the system of two equations coupled through the collision integrals containing intraband and as well interband electron scattering terms.

\section{Electron-electron collision scattering rate}

Solution of this type equations in respect $f_\alpha ({\bf k})$ taking in account the Umklapp processes  is difficult problem. Even for a single band metal with centrum of inversion and spherical Fermi surface 
an analytic solution can be found only by application of variational procedure \cite{Ziman}. In the absence of Umklapp processes,
$ {\bf k}^\prime={\bf k}+{\bf k}_{2}-{\bf k}^{\prime\prime}$,
one can solve this problem in the same manner as it was done 
in the paper \cite{Pal2012} for the two-band metal with centrum of inversion.

The left-hand-side of  equations (\ref{k}) and (\ref{l}) behave like delta-functions
near the Fermi surface of  the corresponding band. Thus, 
following usual linearisation procedure one can search the solution for deviation of distribution function from equilibrium as
\begin{eqnarray}
\delta f_{\alpha}({\bf k})= f_{\alpha}({\bf k})- n(\varepsilon_\alpha)=c_\alpha({\bf v}_{\alpha}{\bf E}) \frac{\partial n({\varepsilon}_\alpha)}{\partial \varepsilon_\alpha},\\
\delta f_{\alpha}({\bf k}_2)= f_{\alpha}({\bf k}_2)- n({\varepsilon}_{2\alpha})=c_\alpha({\bf v}_{2\alpha}{\bf E}) \frac{\partial n({\varepsilon}_{2\alpha})}{\partial \varepsilon_{2\alpha}},
\end{eqnarray} 
etc. Then, multiplying the equation (\ref{k}) and equation (\ref{l}) on ${\bf v}_{+}$ and ${\bf v}_{-}$ correspondingly and integrating over ${\bf k}$
we shall come to the system of two linear algebraic equations for the coefficients $c_+$ and $c_-$.

However, to establish the temperature dependence of the electron-electron relaxation time it is unnecessary to reproduce
 these cumbersome calculations. For this purpose  it is enough to keep in the Eqs. (\ref{k}) and (\ref{l}) only the terms with 
$\delta f_{\alpha}({\bf k})= f_{\alpha}({\bf k})- n(\varepsilon_\alpha)$ neglecting other terms proportional to  $\delta f_{\alpha}({\bf k}_2)$,
etc.  Of course, this trick by no means presents  the solution of kinetic equations. But due to similar energy dependence of all sub-integral terms
the complete calculation yields the same temperature dependence of electron-electron relaxation as it originates from this incomplete treatment.
Thus, ignoring the temperature independent terms due to electron-impurity scattering, we obtain:
\begin{widetext}
\begin{eqnarray}
({\bf v}_{+}{\bf E}) \frac{\partial n({\varepsilon}_+)}{\partial \varepsilon_+}
=-2\pi \delta f_{+}({\bf k})\int \frac{d^3{\bf k}^{\prime\prime}}{(2\pi)^3}\frac{d^3{\bf k}_2}{(2\pi)^3}~~~~~~~~~~~~~~~~~~~~~~~~~~~~~~~~~~~~~
\nonumber\\
\times \left\{W_1 \left [O_{+\nu}({\bf k}, {\bf k}^\prime)) O_{\nu +}({\bf k}^\prime, {\bf k}))
O_{\xi\zeta}({\bf k}_2, {\bf k}^{\prime\prime})
O_{\zeta\xi}({\bf k}^{\prime\prime}, {\bf k}_2)\right ]
+W_2\left   [O_{+\nu}({\bf k}, {\bf k}^\prime)) O_{\nu \xi}({\bf k}^\prime, {\bf k}_2))
O_{\xi\zeta}({\bf k}_2, {\bf k}^{\prime\prime})
O_{\zeta +}({\bf k}^{\prime\prime}, {\bf k}) \right ]\right \}
\times\nonumber\\
~\nonumber\\
\times \left\{  n(\varepsilon_\nu^\prime)(1-n(\varepsilon_{2\xi}))n(\varepsilon_\zeta^{\prime\prime})\right.
+\left.
(1- n(\varepsilon_\nu^\prime)) n(\varepsilon_{2\xi})(1-n(\varepsilon_\zeta^{\prime\prime}) )\right\}
\delta(\varepsilon_\nu^\prime-\varepsilon_{+} -\varepsilon_{2\xi} +\varepsilon_\zeta^{\prime\prime}),~~~~~\\
~\nonumber\\
({\bf v}_{-}{\bf E}) \frac{\partial n({\varepsilon}_-)}{\partial \varepsilon_-}
=-2\pi \delta f_{-}({\bf k})\int\frac{d^3{\bf k}^{\prime\prime}}{(2\pi)^3}\frac{d^3{\bf k}_2}{(2\pi)^3}
 \times \{ +\longrightarrow -\}. ~~~~~~~~~~~~~~~~~~~~~~~~~~~
\end{eqnarray}
\end{widetext}

To take into account the energy conservation one needs to transform the integration over momenta to the integration over energies.
Even for single band metal with centrum inversion one can perform this procedure analytically only in the case of almost spherical shape of the Fermi surface. For a metal  without inversion center with the isotropic spectrum
\begin{equation}
\varepsilon_\pm({\bf k})=\frac{k^2}{2m}\pm\gamma k
\end{equation}
following the procedure developed in the paper  \cite{Khalat} and then reproduced in \cite{Baym1978} in a somewhat different manner,
 we obtain in neglect the terms of the order $\gamma k_F/\varepsilon_F$
\begin{equation}
d^3{\bf k}^{\prime\prime}d^3{\bf k}_2= m^{3}
\frac{\sin\theta
d\theta d\phi d\phi_2}{2\cos(\theta/2)}\left [ 1+{\cal O}\left (\frac{\gamma k_F}{\varepsilon_F}  \right)\right]d\varepsilon^{\prime\prime}_\zeta d\varepsilon_{2\xi}d\varepsilon^\prime_\nu.
\end{equation}
Here $\theta$ is the angle between ${\bf k}$ and ${\bf k}_2 $, $\phi$ is the azimuthal angle of ${\bf k}_2 $ around direction ${\bf k}$, and $\phi_2$ is the angle between the planes 
$({\bf k},{\bf k}_2) $ and $({\bf k}^\prime,{\bf k}^{\prime\prime}) $. 
With the same accuracy one can expect the factors
$O_{\alpha\beta}({\bf k},{\bf k}^\prime)$ as the function dependent only from the angles between the vectors ${\bf k},{\bf k}_{2},{\bf k}^\prime,{\bf k}^{\prime\prime} $.

The integration over $\varepsilon^\prime_\nu$ is reduced to replacement $
\varepsilon^\prime_\nu=\varepsilon_++\varepsilon_{2\xi}-\varepsilon^{\prime\prime}_\zeta.
$ Then performing integration over $\varepsilon^{\prime\prime}_\zeta $ and $\varepsilon_{2\xi}$ we obtain
\begin{eqnarray}
({\bf v}_{+}{\bf E}) \frac{\partial n({\varepsilon}_+)}{\partial \varepsilon_+}
=-m^3
[(\pi T)^2+(\varepsilon_+-\mu)^2] I_+\delta f_{+}({\bf k})~~
\\
({\bf v}_{-}{\bf E}) \frac{\partial n({\varepsilon}_-)}{\partial \varepsilon_-}
=-m^3[(\pi T)^2+(\varepsilon_--\mu)^2] I_-\delta f_{-}({\bf k}),~~
\end{eqnarray}
where
\begin{eqnarray}
I_+=\int\frac{\sin\theta
d\theta d\phi d\phi_2}{2(2\pi)^5\cos(\theta/2)}
~~~~~~~~~~~~~~~~~~~~~
\nonumber\\
\times \left\{W_1 \left [O_{+\nu}({\bf k}, {\bf k}^\prime)) O_{\nu +}({\bf k}^\prime, {\bf k}))
O_{\xi\zeta}({\bf k}_2, {\bf k}^{\prime\prime})
O_{\zeta\xi}({\bf k}^{\prime\prime}, {\bf k}_2)\right ]\right.\nonumber\\
+\left.W_2\left   [O_{+\nu}({\bf k}, {\bf k}^\prime)) O_{\nu \xi}({\bf k}^\prime, {\bf k}_2))
O_{\xi\zeta}({\bf k}_2, {\bf k}^{\prime\prime})
O_{\zeta +}({\bf k}^{\prime\prime}, {\bf k}) \right ]\right \}~~
\end{eqnarray}
and $I_-$     is obtained from $I_+$ by substitution $+\to -$.
Substituting $\delta f_{+}({\bf k})$ and $\delta f_{-}({\bf k})$ to the expression for current
\begin{eqnarray}
 {\bf j}=e^2\int \frac{d^3k}{(2\pi)^3} \left\{
{\bf v}_+\delta f_{+}({\bf k})
+
{\bf v}_-\delta f_{-}({\bf k})\right \}
\label{current}
\end{eqnarray}
and performing the integration we come to
\begin{equation}
{\bf j}=\frac{e^2v_F^2}{3\pi^2 m^3T^2}\left\{\frac{N_{0+}}{I_+}+\frac{N_{0-}}{I_-}\right\}{\bf E},
\end{equation}
where $N_{0\pm}=\frac{mk_{F\pm}}{2\pi^2}$ is the density of states in the $\pm$ bands.   
Thus, if the constraints made during derivation are fulfilled 
 the electron-electron scattering rate and the low temperature resistivity in metals without inversion centre have usual temperature behaviour given by Eqs.(\ref {2}), (\ref{1}).

The mentioned above more complete treatment taking into account all the terms in the equations 
  (\ref{k})-(\ref{56}) including scattering on impurities 
  does not change the  temperature dependence of conductivity.

\section{Conclusion}

There was shown that in the metals without inversion center the electron-electron collisions create the same contribution to the low temperature dependence of resistance as in the ordinary metals without space parity violation. The previous calculations devoted to the same problem \cite{Mineev2018,Mineev2019} have lead to the wrong result  Eq.(\ref{11}) due to using of the  incorrect formulas  for the dispersion laws of electrons.

The presented derivation is valid when the band energy splitting exceeds the  rate of electron-impurity scatterings Eq.(\ref{4}) and at the same time it  is smaller than the energy Fermi ( see Eq.(\ref{5})). Under these conditions the electron-electron scattering causes the usual quadratic temperature dependence of resistivity at low temperatures and does not bring an additional contribution in the metal residual resistivity at $T=0$.

As in ordinary metals  without space parity violation the quadratic temperature dependence determined by electron-electron collisions  can be lost due to the essential anisotropy of the Fermi surface.

\end{document}